\documentclass[]{aa}
\usepackage{epsfig}
\def\simlt{\lower.5ex\hbox{$\; \buildrel < \over \sim \;$}}
\def\simgt{\lower.5ex\hbox{$\; \buildrel > \over \sim \;$}}
\def\H2{\element[][][][2]{H}}

\begin{document}

   \msnr{H2063}

   \thesaurus{08     % A&A Sect. 9: ISM
              (09.03.1; % ISM: clouds
               09.13.1; % ISM: magnetic fields
               09.19.1; % ISM: structure
               02.20.1; % Turbulence
               02.23.1  % Waves
               )}

   \title{Models of Wave Supported Clumps 
       in Giant Molecular Clouds}

   \author{R.F. Coker, J.G.L. Rae, \and T.W. Hartquist}

   \offprints{R.F. Coker}

   \institute{Department of Physics and Astronomy, \\ University of Leeds, 
              Leeds LS2 9JT  UK \\
              email: robc@ast.leeds.ac.uk,jglr@ast.leeds.ac.uk,twh@ast.leeds.ac.uk
             }

   \date{Received 2000; accepted 2000}

   \titlerunning{Turbulent Wave Support of Clumps}
   \authorrunning{Coker, Rae, \& Hartquist}

   \maketitle

   \begin{abstract}

We present plane-parallel equilibrium models of molecular clumps that are supported
by Alfv\'en waves damped by the linear process of ion-neutral friction.
We used a WKB approximation to treat the inward propagation of waves and 
adopted a realistic ionization structure influenced by dissociation and
ionization due to photons of external origin.
The model clumps are larger and less centrally condensed than those
obtained for an assumed ionization structure, used in some previous studies,
that is more appropriate for dark regions.

   \end{abstract}
%________________________________________________________________

\section{Introduction}

Giant molecular cloud complexes (GMCs), the birth places of stars,
are typically many tens of parsecs in linear extent and have masses
from $10^4$ to $10^6 M_{\sun}$ and temperatures of 10--30 K (see
\cite{HCRRW98} for a recent review).
Observations of CO emission from GMCs (\cite{BT80}; \cite{WBS95})
show them to be composed of many smaller clumps that are a few parsecs
in extent and contain $\simlt 10^3 M_{\sun}$.  

The widths of CO emission lines originating in individual clumps
are supersonic and have been attributed to the presence of Alfv\'en waves
having subAlfv\'enic velocity amplitudes (\cite{AM75}).  The
Alfv\'en waves contribute to the support of a clump along the direction
of the large-scale magnetic field; the damping of the waves affects the degree
of support that they provide.  An important and well understood mechanism
for the damping of linear Alfv\'en waves in a partially ionized medium is
that due to ion-neutral friction which depends on the ionization structure
(\cite{KP69}).  Ruffle et al. (1998, hereafter R98) and Hartquist et al. (1993)
have emphasised that the dependence of the ionization structure
on total visual extinction, $A_\mathrm{V}$, should greatly influence the density
profiles of clumps if Alfv\'en waves contribute to their support.  To quantify
the assertion of Hartquist et al. (1993) and R98, we present in this
paper models of plane-parallel, wave-supported GMC clumps like those identified
in the work of Williams et al. (1995), who made a detailed
analysis of the CO maps of the Rosette Molecular Cloud (RMC), identifying
more than 70 clumps.  The models that
we have constructed are for RMC-type clumps in equilibrium, a restriction justified
by the fact that clear spectral signatures of collapse have been found only
when much smaller scale features have been resolved (see, e.g., \cite{HCRRW98}).

We have adopted a WKB description of the wave propagation as did Martin et al. (1997)
in their work on wave-supported clumps.  Their work differs substantially from ours
in that they used an ionization structure appropriate for dark regions.  Also, we have
considered inwardly rather than outwardly propagating waves, as many of the clumps
mapped by Williams et al. (1995) do not contain detected stars and
may have no internal means of generating waves.  Indeed, the waves may be produced
at the surface of a clump by its interaction with an interclump medium.

Other authors have addressed the importance of photoabsorption for the effects
that the ionization structure will have upon a clump's dynamics.  These authors
have been concerned primarily with dense cores and/or envelopes around them;
cores are much smaller-scale objects than the clumps identified in Williams et al. (1995).
McKee (1989) addressed the possibility that collapse in a system of dense
cores is a self-regulating process due to the ionization of metals such as Magnesium and Sodium
by photons emitted by stars formed in the collapse; he was concerned with infall due to
ambipolar diffusion of a large-scale magnetic field.  Ciolek \& Mouschovias (1995)
have shown that the large-scale magnetic field can support a photoionized envelope
around a dense core for a time that is very long compared to the ambipolar diffusion
timescale in the center of the dense core.  In contrast to McKee (1989) and 
Ciolek \& Mouschovias (1995), Myers \& Lazarian (1998) addressed the effect
of photoabsorption on support by waves rather than by the large-scale magnetic
field.  They stressed that observed infall of dense core envelopes is slower than
that expected due to gravitational free-fall and more rapid than collapse due to the
reduction by ambipolar diffusion of support by an ordered large-scale magnetic field.
They considered collapse of material supported primarily by
waves and subjected to an external radiation field.  While they
made clear comments about the importance of the $A_\mathrm{V}$ dependence of the ionization structure for their
model, they did not perform any detailed calculations in which a realistic dependence
of the ionization fraction on $A_\mathrm{V}$ was used.

Several sets of authors have considered
nonlinear effects in the dissipation of waves supporting a clump.  Gammie \& 
Ostriker (1996)
investigated models of plane-parallel clumps and from their
``1 2/3-dimensional'' models found dissipation times due to nonlinear effects to be
longer than the Alfv\'en crossing times for a fairly large range of parameters.
The three dimensional investigations of Mac Low et al. (1998) and 
Stone et al. (1998) suggest the more restrictive condition that the
angular frequency of the longest waves be no more than a few times
$2\pi/t_\mathrm{A}$ (where $t_\mathrm{A}$ is the Alfv\'en crossing time) in order for the
dissipation timescale due to nonlinear damping to be roughly the Alfv\'en crossing time
or more.  In most cases addressed in this paper we have restricted our attention to
such angular frequencies so that we are justified to lowest order in focusing
on only the damping due to ion-neutral friction.  It should be noted
that the above three dimensional studies of nonlinear effects concerned homogeneous
turbulence and did not include ion-neutral damping for a realistic ionization structure.
If we are correct in supposing that the waves in clumps are driven externally,
then the turbulence is not homogeneous 
and its nature depends on both the viscous scale set by ion-neutral damping
and the exact boundary conditions.  The effects of nonlinear damping 
and multiple dimensions will be considered in
subsequent work.

In Sect.
2 we present the equations for the wave energy, the static equilibrium clump structure,
and the gravitational field.  In Sect. 3 we give a description of the calculations of the
ionization structure for various values of the clump density and $A_\mathrm{V}$ while Sect. 4
contains details of the models considered here.  Finally, in Sect. 5, we present
conclusions.

%__________________________________________________________________

\section{Equations of Wave Propagation and Static Equilibrium}

We consider plane-parallel clumps with $z = 0$ corresponding to the clump midplane and
$z = z_\mathrm{b}$ (with $z_\mathrm{b}$ defined as positive) corresponding to boundary between
the clump and the interclump medium.
The large-scale magnetic field is taken to be $B_0 \hat{z}$ with $\hat{z}$ normal to
the surface of a plane-parallel clump.  We study
waves of angular frequency $\omega$ propagating from $z=z_\mathrm{b}$ in the $-\hat{z}$
direction.  

The ion
velocity can be expressed as
\begin{equation}\label{eq:vieq}
v_\mathrm{i} = V e^{-i \int{k_\mathrm{r} {\rm d}z}} \;,
\end{equation}
where $V$ is defined below and $k_\mathrm{r}$ is the real component of the complex wave vector.
Hartquist \& Morfill (1984) used a two-fluid
treatment to examine a related problem and showed that for an
inwardly propagating linear Alfv\'en wave
\begin{equation}\label{eq:conseq}
{{\rm d}\over{\rm d}z} \left(k_\mathrm{r} v_\mathrm{i} v_\mathrm{i}^*\right) = {{V_2}\over{v_{A_\mathrm{i}}^2}} v_\mathrm{i} v_\mathrm{i}^* \;,
\end{equation}
where $v_\mathrm{i}^*$ is the complex conjugate of the ion velocity perturbation and
\begin{equation}\label{eq:V2eq}
V_2 \equiv { {\nu_0 \omega^3 \rho_\mathrm{n}^2} \over {\nu_0^2 \rho_\mathrm{i}^2 + \omega^2 \rho_\mathrm{n}^2} } \;,
\end{equation}
where $\rho_\mathrm{i}$ is ion mass density and $\rho_\mathrm{n}$ is the neutral mass density.
The ion Alfv\'en velocity is given by
\begin{equation}\label{eq:alfeq}
v_{A_\mathrm{i}}^2 = {{B_0}\over{\sqrt{4 \pi \rho_\mathrm{i}}}} \;.
\end{equation}
The ion-neutral coupling frequency, $\nu_0 \rho_\mathrm{i}$, is such that the momentum transfer
per unit volume per unit time from ions to neutrals is given by $\nu_0 \rho_\mathrm{n} \rho_\mathrm{i} (v_\mathrm{i} - v_\mathrm{n})$
where $v_\mathrm{n}$ is the velocity of the neutrals.  To a reasonably
good approximation, C$^+$ is the dominant ion and we may take 
\begin{equation}\label{eq:nu0}
\nu_0 \rho_\mathrm{i} \simeq 2.1\times10^{-9}\mathrm{sec}^{-1} 
\left({{n_\mathrm{i}}\over{1~\mathrm{cm}^{-3}}}\right)
\end{equation}
where $n_\mathrm{i}$ is the ion number density (\cite{O61}).
If the dominant ion species are very massive, as occurs at large $A_\mathrm{V}$ and densities,
the constant would approach $2.3\times10^{-9}$s$^{-1}$.  However, in this work we
ignore this dependence and use $12~m_\mathrm{H}$ as the mass per ion.

If $V$ in Eq.~\ref{eq:vieq} is written as $e^{\int{k_\mathrm{i} {\rm d}z}}$,
where $k_\mathrm{i}$ is the imaginary component of the wave vector, and the linearized version
of the induction equation (cf. Eq. 3 of Hartquist \& Morfill 1984) is used, Eq.~\ref{eq:conseq}
yields
\begin{equation}\label{eq:conseq2}
{{\rm d}\over{\rm d}z} \left({{k_\mathrm{r}}\over{k_\mathrm{r}^2 + k_\mathrm{i}^2}} b b^*\right) = {{V_2}\over{v_{A_\mathrm{i}}^2}} 
{{b b^*}\over{k_\mathrm{r}^2 + k_\mathrm{i}^2}} \;,
\end{equation}
where $b$ is the perturbation magnetic field and $b^*$ is its complex
conjugate.  In the WKB approximation, d$^2V/$d$z^2$ is taken
to be equal to zero, which is equivalent to assuming
\begin{equation}\label{eq:ass}
k_\mathrm{i}^2 + {\rm d}k_\mathrm{i}/{\rm d}z \ll k_\mathrm{r}^2
\end{equation}       
Then
it follows (cf. Eq. 7b of Hartquist \& Morfill 1984) that in the WKB approximation
\begin{equation}\label{eq:WKBeq}
k_\mathrm{r}^2 \simeq {{\omega^2 + V_1} \over{ v_{A_\mathrm{i}}^2}} \;,
\end{equation}
with 
\begin{equation}\label{eq:V1eq2}
V_1 \equiv {{ \nu_0^2 \omega^2 \rho_\mathrm{n} \rho_\mathrm{i}} \over {\nu_0^2 \rho_\mathrm{i}^2 + \omega^2 \rho_\mathrm{n}^2} }\;.
\end{equation}
As is consistent with the WKB approximation, we assume $k_\mathrm{i}^2 \ll k_\mathrm{r}^2$ and substitute 
Eq.~\ref{eq:WKBeq} into Eq.~\ref{eq:conseq2} to find
\begin{equation}\label{eq:conseq3}
{{\rm d}\over{\rm d}z} \left({{v_{A_\mathrm{i}} U}\over{\sqrt{\omega^2 + V_1}}} \right) = 
{{V_2}\over{\omega^2 + V_1}} U \;,
\end{equation}
where $U = bb^*/16\pi$ is the time-averaged energy density of the perturbation
magnetic field.

We solve Eq.~\ref{eq:conseq3} along with the static equilibrium equation
\begin{equation}\label{eq:eqlbrm}
c_s^2 {{ {\rm d} (\rho_\mathrm{n} + \rho_\mathrm{i})}\over{\rm d}z} + {{ {\rm d}U}\over{ {\rm d}z}} =
-(\rho_\mathrm{n} + \rho_\mathrm{i}) g \;
\end{equation}
(\cite{MHP97})
and the gravitational equation
\begin{equation}\label{eq:grav}
{{ {\rm d}g} \over{\rm d}z}  = 4 \pi G ( \rho_\mathrm{n} + \rho_\mathrm{i}) \;,
\end{equation}
where $c_s$, $g$, and $G$ are the isothermal sound speed, the strength of the
gravitational field, and the gravitational constant, respectively.

We verify the assumption that $k_\mathrm{i}^2 \ll k_\mathrm{r}^2$ {\sl a posteri} by checking that 
\begin{equation}\label{eq:test}
k_\mathrm{i} = {{1}\over{V}}{{{\rm d}V}\over{{\rm d}z}} \simeq {{V_2}\over{v_{A_\mathrm{i}}\sqrt{V_1}}} \ll {{V_1}\over{v_{A_\mathrm{i}}^2}} \;.
\end{equation}
Note that
if Eq.~\ref{eq:test} is satisfied, Eq.~\ref{eq:ass} is as well.

%________________________________________________________________

\section{Calculations of the Ionization Structure}

The ionization structure determined by R98 and presented in their Fig. 1 
was calculated on the assumption that, due to shielding of the CO by itself and by H$_2$, the rate
of CO dissociation by photons of external origin is negligible.  
For a plane-parallel semi-infinite cloud with constant Hydrogen nucleus number density,
$n_\mathrm{H} =10^3~$cm$^{-3}$, we
assume an $A_\mathrm{V}$-dependent dissociation rate that results
in a CO abundance relative to $n_\mathrm{H}$, x(CO), that is
in harmony with the measurements shown in Fig. 6 of van Dishoeck (1998).
For the $n_\mathrm{H} = 10^3~$cm$^{-3}$ model, Table~\ref{tab:rates} gives x(CO) and
the photodissociation rate as a function of $A_\mathrm{V}$.  Note that the total abundance
of carbon nuclei relative to $n_\mathrm{H}$ is fixed at $10^{-4}$.  
In the work reported here, we used an ionization fraction that depends
on both $A_\mathrm{V}$ and $n_\mathrm{H}$. 
Using the $A_\mathrm{V}$-dependent CO photodissociation rate from Table~\ref{tab:rates},
we calculated the fractional ionization as a function of $A_\mathrm{V}$ 
for $n_\mathrm{H} = 3\times10^2, 3\times10^3, $~and~$ 10^4~$cm$^{-3}$ at various $A_\mathrm{V}$ values.  A bilinear
interpolation in $A_\mathrm{V}$ and $n_\mathrm{H}$ is used to find the actual ionization
fraction used for a given point in the clump.  For densities above $10^4~$cm$^{-3}$
we assume that the total ionization fraction, $\xi \equiv \rho_\mathrm{i}/\rho_\mathrm{n}$, goes as $n_\mathrm{H}^{-1/2}$,
as expected in a dark region.  Note that in all models, $A_\mathrm{V}$ is always
greater than 4 whenever $n_\mathrm{H} > 10^4~$cm$^{-3}$.

Since the depletions in RMC-type clumps are very uncertain,
we present results for both depletion cases given in R98.  As discussed therein
(also see Shalabiea \& Greenberg 1995), case A abundances resemble those seen
in dark cores with $A_\mathrm{V}\simgt 5$ while case B, with higher fractional abundances of lower
ionization potential elements, is more appropriate for more diffuse clouds.

\begin{table}
  \caption{Assumed CO Photodissociation Rates}
  \label{tab:rates}
  \begin{tabular}{lll}
    \hline
           &                     &                       \\ [-11pt]
 $A_\mathrm{V}$     &Rate (sec$^{-1}$)    &x(CO)                  \\
           &                     &                       \\ [-11pt]
    \hline
           &                     &                       \\ [-10pt]
 0.5       &$7.685\times10^{-13}$&$9.132\times10^{-7}$ \\
 1.0       &$4.279\times10^{-13}$&$2.248\times10^{-6}$ \\
 1.5       &$2.132\times10^{-13}$&$5.248\times10^{-6}$ \\
 1.75      &$1.440\times10^{-13}$&$7.538\times10^{-6}$ \\
 2.0       &$7.465\times10^{-14}$&$1.226\times10^{-5}$ \\
 2.25      &$4.864\times10^{-14}$&$1.503\times10^{-5}$ \\
 2.5       &$2.245\times10^{-14}$&$2.166\times10^{-5}$ \\
 2.75      &$1.405\times10^{-14}$&$2.381\times10^{-5}$ \\
 3.0       &$4.291\times10^{-15}$&$3.621\times10^{-5}$ \\
 3.25      &$4.112\times10^{-15}$&$3.620\times10^{-5}$ \\
 3.5       &$3.669\times10^{-16}$&$6.336\times10^{-5}$ \\
 3.75      &$1.200\times10^{-16}$&$8.976\times10^{-5}$ \\
 4.0       &$       0           $&$9.767\times10^{-5}$ \\
     \hline
  \end{tabular}
\end{table}

%________________________________________________________________

\section{Details of the Models}

Many RMC-type clumps are not bound by their own gravity and must be confined
by interclump media
(\cite{BM92}).
We shall assume that an interclump medium is sufficiently tenuous that
it does not shield the clump from the standard interstellar background
radiation field used in the calculation of $\xi$ so that $A_\mathrm{V} = 0$ at
$z = z_\mathrm{b}$.  The material on either side of the interface between a clump
and the interclump medium is in two distinct phases, and we may assume
that at its outer boundary a clump has a 
substantial density; we take $n(\H2) = n_\mathrm{H} / 2$ everywhere in the clump
and at $z = z_\mathrm{b}$ set $n(\H2) = n_\mathrm{b}$. 
Waves may exist in the interclump medium and be partially
transmitted into the clump, or, as mentioned earlier, may be generated
near the interface by the interaction between the clump and the interclump medium.
Consequently, we assume that the magnitude of the amplitude of the perturbation
magnetic field, $\delta B = \sqrt{bb^*}$, is, at $z = z_\mathrm{b}$, a substantial fraction, $f_\mathrm{b}$,
of the large-scale field, $B_0$.

In all models we have taken the mean mass per neutral particle, $\mu_\mathrm{n}$, to be
2.3 amu, corresponding to $14\%$ of the neutral particles being He and $86\%$ being $\H2$.
As is consistent with data given by Savage \& Mathis (1979), we have assumed 
\begin{equation}
A_\mathrm{V} = {{N_\mathrm{H}}\over{1.9\times10^{21}{\rm cm}^{-2}}} \;,
\end{equation} where
$N_\mathrm{H}$ is the column density of Hydrogen nuclei.  Also, the temperature throughout a clump was
taken to be 20 K.

For a given setup, an initial boundary value of $g(z=z_\mathrm{b})\equiv g_\mathrm{b}$, was selected and 
Eqs.~\ref{eq:conseq3},~\ref{eq:eqlbrm},
and ~\ref{eq:grav} were numerically integrated using an adaptive Gear
algorithm (\cite{G71}).  The value of $g_\mathrm{b}$ was changed by iteration
until the inner boundary condition $g(z=0)=0$ was satisfied.  
Note that as the total visual extinction (or,
equivalently for a plane-parallel cloud, the column density) for a cloud
is increased, the clump reaches a maximum size, $z_\mathrm{max}$, and
starts shrinking as a larger and larger thermal pressure (and
therefore density) is required to balance gravity.  Thus, as
long as $z_\mathrm{b} < z_\mathrm{max}$, there are
two values of $g_\mathrm{b}$ for each setup that satisfy the boundary conditions.
One solution has a relatively flat density profile and a small total
visual extinction while the
other solution is more centrally condensed.  We deal here exclusively
with the latter solutions.

We have considered other models but present full results only for
models which have a total edge-to-center visual extinction of
5 magnitudes since, as discussed in R98, it is in the region of $A_\mathrm{V}$ of a few that 
clumps appear to begin to contain detected stars, while many dense cores may
have $A_\mathrm{V} \simlt 5$ (\cite{M89}).  For our canonical model (Model 1) we require
a velocity amplitude, $V$, of 2 km sec$^{-1}$ and an Alfv\'en speed, $v_\mathrm{A}$, of
3 km sec$^{-1}$ at $A_\mathrm{V} = 2$.  Since the concentration of CO at $A_\mathrm{V} \simlt 2$ is
very low and measurements of
GMC clump CO profiles have a width of $\sim 2~$ km sec$^{-1}$ 
(\cite{WBS95}),
observations require
such velocities to exist well within the cloud.  Also, we use a wave frequency, $\omega$,
of $2\times10^{-12}$~sec$^{-1}$; this results in relatively strong neutral-ion coupling
while keeping the wavelength of the perturbing wave less than the size of the
cloud.  In other words,
\begin{equation}\label{eq:omega}
\nu_0 \xi \ll \omega \simlt {{2 \pi}\over{\int{\rm d}z/v_\mathrm{A}}} \;.
\end{equation}                                  
Within the above constraints, we find for our canonical model the solution requiring the smallest
value of $B_0$, and, since the velocity amplitude of
a linear Alfv\'en wave is thought to be comparable to but less than the Alfv\'en speed, the largest
value of $f_\mathrm{b}$.  However, throughout the clump, $V \leq v_\mathrm{A}$,
consistent with our assumption of linear Alfv\'en waves.

We present the results of 5 models.  Model 1 is for the above canonical parameters
with R98's depletion case A while Model 2 is for case B.  Model 3 is the same as
Model 1 but with $\omega = 1\times10^{-12}$ to illustrate the effect of a scenario
with roughly maximum ion-neutral coupling.  Model 4 is as Model 1 but
with an ionization profile given by $\xi = 3\times10^{-16} {\rho_\mathrm{n}}^{-1/2}$
(see, e.g., McKee 1989 and Myers \& Lazarian 1998).
Thus, Model 4 is for a dark region surrounded by interclump material.
Finally, for completeness, we present a model with no turbulence; Model 5 is 
identical to Model 1 except with $f_\mathrm{b} = 0$.  A summary of the parameters of
the 5 models is given in Table~\ref{tab:models}.  For all models, $n_\mathrm{b} = 375~\H2$ cm$^{-3}$
and $B_0 = 135~\mu$G.

\begin{table}
  \centering
  \caption{Summary of Parameters used in Models}
  \label{tab:models}
  \begin{tabular}{lllll}
    \hline
           &                     &         &             &              \\ [-11pt]
 Model     &$\omega$ (sec$^{-1}$)& $f_\mathrm{b}$   & $z_\mathrm{b}$ (pc)  & Depletions   \\
           &                     &         &             &              \\ [-11pt]
    \hline &                     &         &             &              \\ [-10pt]
 1         &$2\times10^{-12}$    & 0.436   & 0.455      & A             \\
 2         &$2\times10^{-12}$    & 0.436    & 0.4855     & B             \\
 3         &$1\times10^{-12}$    & 0.436    & 0.495      & A             \\
 4$^{\mathrm{a}}$&$2\times10^{-12}$    & 0.436    & 0.1978     & A             \\
 5         &$2\times10^{-12}$    & 0.0      & 0.082      & A             \\
     \hline
  \end{tabular}
  \begin{list}{}{}
  \item[$^{\mathrm{a}}$] Uses $\xi = 3\times10^{-16} {\rho_\mathrm{n}}^{-1/2}$.
  \end{list}
\end{table}                                 

%________________________________________________________________

\section{Results and Conclusions}

   \begin{figure}
      \epsfxsize=8.8cm \epsfbox{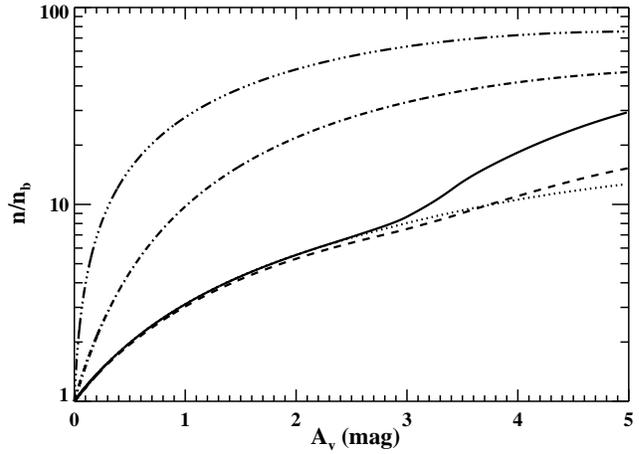} 
      \caption[]{Plot of density (normalized to the density at the outer edge of the
       clump) versus $A_\mathrm{V}$ for the 5 Models.  The solid
       curve is for Model 1, the dotted curve for Model 2, the dashed curve for Model 3,
       the dot-dashed curve for Model 4, and the dash-chain-dot curve for Model 5.}
       \label{fig:rho}
    \end{figure}     

   \begin{figure}
      \epsfxsize=8.8cm \epsfbox{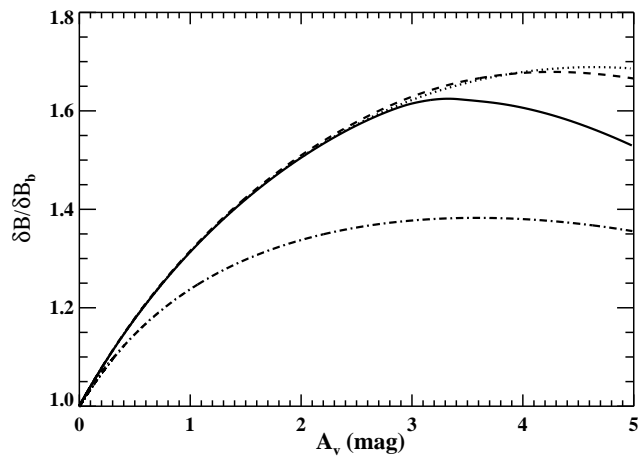} 
      \caption[]{As Fig.~\ref{fig:rho} but for the perturbing magnetic field, $\delta B$.}
      \label{fig:deltaB}
    \end{figure}     

   \begin{figure}
      \epsfxsize=8.8cm \epsfbox{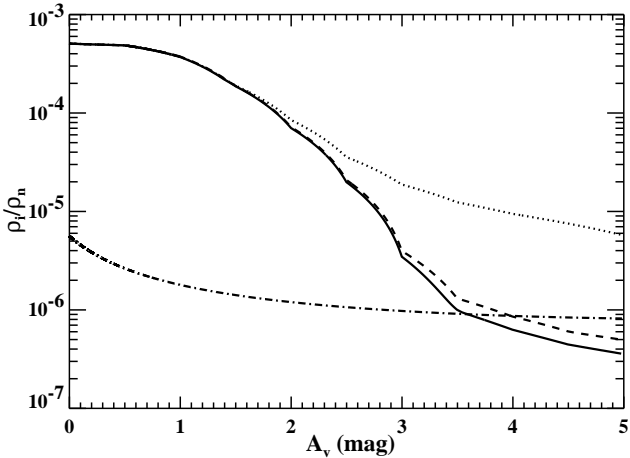} 
      \caption[]{As Fig.~\ref{fig:rho} but for the absolute ion mass fraction, $\xi$.  The
       bumps seen on the curves for Models 1--3 are a result of the interpolation over
       $A_\mathrm{V}$.}
       \label{fig:xi}
    \end{figure}     

   \begin{figure}
      \epsfxsize=8.8cm \epsfbox{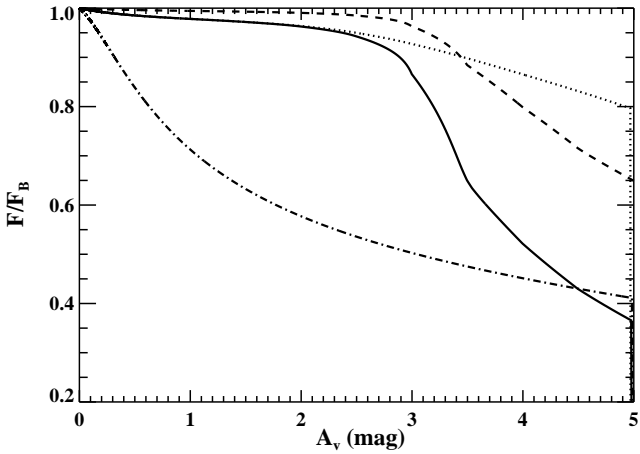} 
      \caption[]{As Fig.~\ref{fig:rho} but for the magnetic energy flux, $F \equiv k_\mathrm{r} U \omega / |k|^2$.}
      \label{fig:flux}
    \end{figure}     

   \begin{figure}
      \epsfxsize=8.8cm \epsfbox{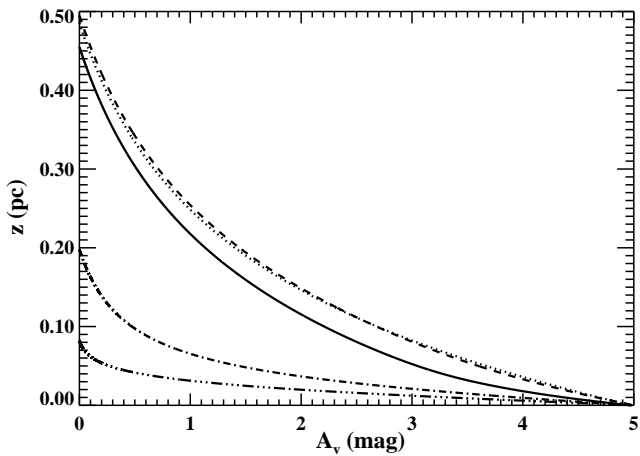} 
      \caption[]{As Fig.~\ref{fig:rho} but for z, the absolute spatial extent of the plane-parallel cloud.}
      \label{fig:size}
    \end{figure}     

In Fig.~\ref{fig:rho} we present density as a function of visual extinction for each of the 5 Models.
It is clear that the newer ionization profiles used in Models 1--3 result
in less condensed, more extended clumps than the profile used in Model 4.  In fact, $n_\mathrm{H}$ is roughly proportional to $1/z$ 
in Models 1--3 while $n_\mathrm{H}$ goes roughly as $1/z^2$ in Model 4.  In Fig.~\ref{fig:deltaB} we show
plots of $\delta B$ versus $A_\mathrm{V}$ for Models 1--4.  Except for Model 4, which has $k_\mathrm{i}/k_\mathrm{r}$ approaching
$0.5$ at the clump boundary so that Eq.~\ref{eq:test} is not satisfactorily satisfied,
the perturbing field obeys flux conservation
near the surface of the clump.  In Model 1, in the central region of the clump 
dissipation is rapid enough that $\delta B$ begins to decrease.  In order to compensate
for the loss of support, the equilibrium solution requires a complementary increase in the density,
as can be seen in Fig.~\ref{fig:rho}.  On the other hand, in Model 4, the turbulence is dissipated
much nearer to the cloud boundary, thus requiring a steeper overall density profile.
The higher ionization fractions of the case B depletions result in very little 
dissipation even in the center of the clump for Model 2.
Note that even though
observations (\cite{WBS95}) suggest that the
temperature of RMC-type
clumps is closer to 10 K rather than the 20 K used here, thermal
support is insignificant except in the centre of Model 1, so the effect of a lower
temperature on the models would be merely to enhance slightly any central condensations.        

The ionization profiles used in the Models are shown in Fig.~\ref{fig:xi}.  The ionization profiles
described in Sec. 3 result in $\xi$ for Models 1--3 being more than 50 times greater near the surface
of the clump than in Model 4.  However, in the center of the clump
the ionization fraction drops, resulting in more dissipation.  Again, this leads to clumps
that are overall more diffuse but with small condensed cores.  Clearly, clumps with $A_\mathrm{V} \simgt 5$
will have distinct central condensations with $n/n_\mathrm{b} \simgt 100$ and central fractional ionizations of $\simlt
5\times10^{-7}$.  Though dense cores may be formed during the fairly rapid collapse
(as envisaged by Fielder \& Mouschovias 1993) of more extended objects (i.e. RMC-like clumps)
that become unstable, even in our equilibrium models we find central cores having densities and fractional
ionizations similar to those measured for dense cores and their envelopes (\cite{WBS95}; \cite{WBCMP98}; \cite{BPWM99}).

Fig.~\ref{fig:flux} shows the flux of magnetic energy
through the clumps for Models 1--4.  Near the clump center, Model 1 is nearly thermally supported due to the
dissipation of the turbulence.
The higher ionization fraction for the case B depletions used in Model 2 results in less
dissipation and thus more turbulent support for the clump.  Consequently, as can be
seen in Fig.~\ref{fig:rho}, Model 2 has no central condensation and is more extended.  Unfortunately,
we can only speculate about how the depletions of Sulphur, metals, and some other species
behave in RMC-like clumps (\cite{RHCW99}).  Thus, 
cases A and B are merely representative; as can
be clearly seen in the figures, the clump profiles are very sensitive to the choice of
abundances and the subsequent fractional ionizations.
In addition, compared to Model 1, the stronger ion-neutral coupling in Model 3 results in less dissipation
and subsequently the clump has little central condensation, as expected.  Note however that
for Model 3 the lower limit in Eq.~\ref{eq:omega} is not adequately satisfied.

Fig.~\ref{fig:flux} also shows the effect of external wave generation.  If the fractional ionization
is too low, as in Model 4, dissipation occurs close to the surface of the clump.
Conversely, if the fractional ionization is too large, as in Model 2, significant
dissipation occurs only at the clump's very centre.  Both extremes produce density profiles
which lack a central condensation.  Note that if our externally generated wave model
is correct, one should not see turbulence within a condensed core if there is no
turbulence in its surrounding envelope.

In Fig.~\ref{fig:size} we present curves which map the visual extinction to the spatial extent of
the clumps.  Clouds with larger extents which match the observed 2--3 pc size of RMC-type
clumps (\cite{WBS95}) cannot be reproduced within the constraints given
in Sect. 4.  However, observations generally measure the largest linear extent of
a clump.   Thus, since the waves only support the model clumps parallel to the large-scale
field, it is not surprising that the model sizes given here are less than the observed sizes.

Similarly, the models require high boundary
densities and magnetic field strengths in order for the Alfv\'en speed and wave velocity
amplitude at visual extinctions
where CO is abundant to be large enough to be compatible with observed linewidths.
For Model 1, $n_\mathrm{b}  = 375~\H2$ cm$^{-3}$.  This is rather higher than the typical value of
$n(\H2) = 220$ cm$^{-3}$ given by Williams et al. (1995) for RMC-type clumps but,
given the uncertainties, it is within a reasonable range of
the Williams et al. (1995) value.  For Model 1, $B_0 = 135~\mu$G,
significantly higher than the value of $30\, \mu$G suggested
by observations (\cite{H87}) and expected from robust theoretical arguments
(\cite{M87}).  In order to determine
whether the values of $n_\mathrm{b}$ and $B_0$ could be lower and still allow model
properties to be consistent with observed linewidths, we constructed models for 
clumps with total edge-to-center extinctions of 3 magnitudes.  The model giving $V = 2$ km sec$^{-1}$
and $v_\mathrm{A} = 3$ km sec$^{-1}$ at $A_\mathrm{V} = 2$ had $n_\mathrm{b} = 325~\H2$ cm$^{-3}$, $B_0 = 105\,\mu$G, 
$f_\mathrm{b} = 0.49$, and $z_\mathrm{max} = 0.565$ pc; although $n_\mathrm{b}$ and $B_0$ were
smaller and $z_{max}$ larger, the agreement with observations is nonetheless poor.

Thus, the next step in the modelling of clumps
in which wave support is important is the inclusion of
wave support in models analogous to the axisymmetric models
of magnetically and thermally suported clumps described in classic papers by
Mouschovias (1976a,1976b).  It is possible that the inclusion of magnetic tension, as well as
pressure, will allow the reduction
of $B_0$ to a value more like that expected and the construction of models of clumps having larger linear extents.

%\begin{acknowledgements}
%      Part of this work was supported by somebody.
%\end{acknowledgements}

\end{document}